\documentclass[plb,reprint]{revtex4-1}
\usepackage{amssymb,amsfonts,amsmath,graphicx}

\usepackage{amssymb,amsmath,epsfig,todonotes}

\newcommand{\cD}{{\cal D}}
\newcommand{\cT}{{\cal T}}
\newcommand{\cV}{{\cal V}}
\newcommand{\IP}{{\mathbb P}}

\newcommand{\cA}{{\mathcal A}}
\newcommand{\cM}{{\mathcal M}}

\begin{document}

\begin{flushright}
 {CERN-TH-2019-046}\\
\end{flushright}

\title{Distinguishing Elliptic Fibrations with AI}



\author{Yang-Hui He}
\affiliation{Department of Mathematics, City, University of London, EC1V0HB, UK,}
\affiliation{Merton College, University of Oxford, OX14JD, UK,}
\affiliation{School of Physics, NanKai University, Tianjin, 300071, China}
\email[]{hey@maths.ox.ac.uk}

\author{Seung-Joo Lee}
\affiliation{CERN, Theory Department,
1 Esplande des Particules, Geneva 23, CH-1211, Switzerland}
\email[]{seung.joo.lee@cern.ch}


\begin{abstract}\vskip 3mm\noindent
We use the latest techniques in machine-learning to study whether from the landscape of Calabi-Yau manifolds one can distinguish elliptically fibred ones.
Using the dataset of complete intersections in products of projective spaces (CICY3 and CICY4, totalling about a million manifolds) as a concrete playground, we find that a relatively simple neural network with forward-feeding multi-layers can very efficiently distinguish the elliptic fibrations, much more so than using the traditional methods of manipulating the defining equations.
We cross-check with control cases to ensure that the AI is not randomly guessing and is indeed identifying an inherent structure.
Our result should prove useful in F-theory and string model building as well as in pure algebraic geometry.
\end{abstract}


\maketitle


\section{Introduction and Summary}

Ever since the birth of string theory its compactifications on compact Calabi-Yau manifolds have been a subject of constant interest to theoretical physicists and algebraic geometers alike. 
Such compactifications provide a well-controlled setup to study supersymmetric effective theories, many physical properties of which can be calculated via algebro-geometric tools without explicit Ricci flat metrics at hand. 
Tremendous efforts to find realistic models of string theory in this setup have been one of the most important aims of string phenomenology; see~\cite{Candelas:1985en} for an earliest attempt.
Furthermore, in line with the so-called swampland program~\cite{Vafa:2005ui}, geometric origins of quantum gravity constraints have recently started to be addressed from the universal properties of general Calabi-Yau manifolds in various string theoretic setups~\cite{Lee:2018urn, Lee:2018spm, Lee:2019tst, Lee:2019xtm, Corvilain:2018lgw, Grimm:2018ohb, Grimm:2018cpv, Marchesano:2019ifh, Font:2019cxq}. 

More generally, non-Ricci-flat manifolds may also lead to supersymmetric compactifications, with an appropriately varying axio-dilaton profile turned on, in the framework of F-theory~\cite{Vafa:1996xn}. This provides arguably the most general geometric approach currently available to study non-perturbative string compactifcations. 
Interestingly, elliptically fibered Calabi-Yau manifolds of one-dimension higher, which are fictitious at least from the Type IIB point of view, can completely specify such compactifications. 
It is therefore of utmost interest to string theorists to better understand elliptic Calabi-Yau manifolds.    

Elliptic fibration structures often bring in a computational power as well. For instance, the first exact MSSM particle content directly from string theory compactifications arose from an elliptically fibred Calabi-Yau threefold, a quotient of the so-called Sch\"on manifold, whose elliptic-fibration structure was what allowed the explicit computation of equivariant cohomology in order to obtain the matter representations \cite{Braun:2005ux,Braun:2005nv,Bouchard:2005ag,Braun:2006ae}.

In the meantime various dualities amongst Calabi-Yau compactifications have been influential in exciting development in enumerative and algebraic geometry~\cite{MS, CK} as well as many other branches of pure mathematics. 
Interestingly, such string dualities oftentimes base on elliptic fibration structures within the internal Calabi-Yau manifolds; see e.g.~\cite{Anderson:2016cdu}. 

While it still remains an open question if Calabi-Yau $n$-folds form a finite set for $n \geq 3$, finiteness of elliptic ones with dimension $n=3$ has been established~\cite{gross, grassi} (see also~\cite{dicerbo} for relevant results for $n=4$ and $5$). 
At the same time, recent investigations of vast Calabi-Yau datasets have observed ubiquity of elliptic fibrations~\cite{Gray:2014fla, Anderson:2017aux, Huang:2018esr, Anderson:2018kwv, Rohsiepe:2005qg, Johnson:2014xpa, Johnson:2016qar, Candelas:2012uu}.
Hence, also in the hope of providing a meaningful measure for the potential paucity of non-ellipticity, there arises a pressing question of distinguishing the elliptic Calabi-Yau manifolds from the non-elliptic. 

In principle, there are methods to determine whether a Calabi-Yau manifold of dimension not bigger than $3$ is an elliptic fibration~\cite{oguiso, pmhwilson}, which were conjectured to work also for a higher-dimensional manifold~\cite{kollar}.
However, 
the computation, as is with any computation in algebraic geometry, could often become rather intense in practice because one needs to first determine the K\"ahler cone of the geometry in question to apply the methods. 
Another source of difficulty is that manipulating high degree polynomials in many variables is a non-linear and complicated matter \cite{NLA}.

In \cite{He:2017aed,He:2017set} a paradigm was proposed to attempt to use artificial intelligence (AI) to bypass expensive algorithms in computational geometry, in particular to study the string landscape and beyond.
It was found that central problems such as computing cohomology of vector bundles appear to be machine-learnable to very high precision. 
Indeed, \cite{He:2017aed,He:2017set,Ruehle:2017mzq,Krefl:2017yox,Carifio:2017bov} brought machine learning to the landscape and there has subsequently been a host of activity successfully addressing various problems in the string landscape using AI techniques and machine-learning \cite{Jejjala:2019kio,Cole:2018emh,Mutter:2018sra,Rudelius:2018yqi,Halverson:2018cio,Klaewer:2018sfl,Demirtas:2018akl,Bull:2018uow,Jinno:2018dek,Cunningham:2018sdj,Wang:2018rkk,Demir:2018iqo,Hashimoto:2018ftp,Carifio:2017nyb,Liu:2017dzi,Carifio:2017bov,Cohen:2017exh,Ruehle:2017mzq,Krefl:2017yox,Bull:2019cij, Wang:2018rkk, Halverson:2019tkf} (cf.~\cite{He:2018jtw} for a pedagogical introduction).
It is therefore natural to ask whether machine-learning techniques can be employed into our present problem of recognizing elliptic fibrations.

The purpose of this letter is to report that this problem of recognizing elliptic fibrations within Calabi-Yau manifolds (and presumably more arbitrary dataset of algebraic varieties), using complete intersection 3-folds and 4-folds within products of projective spaces  as a playground, appear to belong to the class of problems which can be addressed by machine-learning to high precision.
Thus, like computing cohomology of bundles over varieties, distinguishing elliptic fibrations appears to be a pattern recognizable by the likes of a neural network - completely without any knowledge of algebraic geometry or expensive algorithms needed to deterministically address the problem.

The outline of the letter is as follows.
In Section II, we briefly set the scene by reminding the reader of the CICY (complete intersection Calabi-Yau manifolds in products of projective spaces) dataset on which we will focus for concreteness, and of the elliptic fibration strucures within the set.
Then, in Section III, we establish a neural network to efficiently learn and identify which of the set are (not) elliptically fibred, for the 3-fold and 4-fold cases, as well as a contrived control case which gives a good sanity check.
Finally, in Section IV, we end with conclusions and outlook.

\paragraph*{{\bf Acknowledgments}}
We thank B.~Assel, W.~Lerche, K.~Siampos and T.~Weigand for helpful discussions and A.~Grassi for comments.
YHH would like to thank STFC for grant ST/J00037X/1.
The work of SJL is supported by the Korean Research Foundation (KRF) through the CERN-Korea Fellowship program. 



\section{CICYs and Elliptic Fibrations}\label{CICY-EF}
As stated in the introduction, we will focus on the so-called CICY dataset for concreteness.
These are smooth Calabi-Yau manifolds embedded as complete intersection in products of projective spaces; we start by briefly recalling their construction. 

Let us first consider a complete intersection ${\mathcal M}$ of $K$ polynomials, $p_{\alpha=1, \cdots, K}$, in the product ambient space, $\mathcal A= \IP^{n_1} \times \cdots \times \IP^{n_m}$. By construction the complex dimension $n$ of $\mathcal M$ is given as 
$n=\sum\limits_{r=1}^m n_r - K\,.$
Denoting the degrees of the defining polynomials $p_\alpha$ in the $r$-th projective spaces by $a^{r}_\alpha$, we can describe a family of such geometries by a {\it configuration} matrix $M$ of the form, 
\begin{eqnarray}\label{conf}
M:=\left[ \, \mathbf n \, | \,  \{\mathbf a_\alpha\} \, \right] 
= \def\arraystretch{1.15}\left[\begin{array}{c|ccc} 
\IP^{n_1} &\, a^{1}_{1} & \cdots &a^{1}_{K} \\ 
\vdots &  \vdots &\ddots&\vdots\\
\IP^{n_m} &\, a^{m}_{1} & \cdots& a^{m}_{K}\\ 
\end{array}\right]  \ , 
\end{eqnarray}
with non-negative integer entries $a^r_\alpha$ (c.f. see~\cite{Anderson:2015iia} for a generalization to include ``negative degrees'', dubbed as gCICYs, and also~\cite{Anderson:2015yzz} for an example study of fibration structures thereof). Such a complete intersection is a Calabi-Yau $n$-fold when 
$\sum\limits_{\alpha=1}^K a_\alpha^r = n_r+1$ 
for every $r=1,\dots, m$. 

In this letter, we will be concerned with CICY 3-folds and 4-folds, for which $7,890$ and $921,497$ configurations were classified, respectively, in~\cite{cicy} and~\cite{Gray:2013mja}. In particular, we will focus on elliptic fibration structures that those CICYs may admit. As first studied systematically in~\cite{Gray:2014fla}, there is a simple combinatorial method to find elliptic fibrations in terms of their configuration matrices~\eqref{conf}. That is, a CICY $\cM$ has an obvious elliptic fibration (OEF) if, via row and column permutations, its configuration $M$ can be put in the form,
\begin{eqnarray}\label{conf-split}
\def\arraystretch{1.15}\left[\begin{array}{c|ccc} 
\cA_1 &\, 0 & \mathcal F \\ 
\cA_2 &\, \mathcal B & \mathcal T\\ 
\end{array}\right]  \ , 
\end{eqnarray}
with the sub-configuration $F:=\left[\cA_1\,|\, \mathcal F \right]$ describing a subvariety of dimension one. Here, $\cA_1$ and $\cA_2$ are products of $m_1$ and $m_2:=m-m_1$ projective spaces, respectively, while $\mathcal F$, $\mathcal B$ and $\mathcal T$ are sub-block matrices. 

While not every elliptic fibration is necessarily an OEF, upon a classification of fibrations~\cite{Anderson:2017aux}, it has been observed that all elliptic CICY $3$-folds do admit an OEF. To be specific, $99.33 \%$ (all but $53$) of the $7,868$ CICY $3$-fold configurations with genuine $SU(3)$ holonomy admit an OEF structure, where the $53$ not only lack an OEF but also cannot be elliptic at all. 
This classification of elliptic fibrations (EFs) within the CICY3 dataset, as opposed to OEFs, was achieved based on the following topological criteria conjectured in~\cite{kollar}:

\noindent 
{\it A Calabi-Yau $n$-fold $\cM$ is elliptic iff there exists a $(1,1)$-class $D \in H^2(\cM,\mathbb Q)$ such that} (a) {\it $D \cdot C \geq 0$ for every algebraic curve $C$}; (b) {\it $D^{n-1} \neq 0$}; (c) {\it $D^n =0$}. 

\noindent 
This has been proven for the $3$-fold case subject to the constraints that $D$ is effective or $D \cdot c_2(\cM) \neq 0$~\cite{oguiso, pmhwilson}.

Similarly, OEFs of CICY $4$-folds were classified~\cite{Gray:2014fla}, resulting in the observation that $99.95 \%$ (all but $462$) of the $905,684$ $4$-fold configurations with $SU(4)$ holonomy admit an OEF. Although an analogous classification of EFs have not been undertaken (and hence, it is a priori not clear whether or not there exists an elliptic CICY $4$-fold lacking an OEF), we will nevertheless use the existence of an OEF as a measure for the ellipticity of a CICY $4$-fold when training AI. 
Table \ref{t:ell} summarizes relevant statistics and characteristic features~\cite{Anderson:2017aux, Gray:2014fla} on OEF and EF for the CICY3 and CICY4 cases.

\begin{table}[h!t!]
\begin{tabular}{|c|c|c|c|c|}\hline
	& Total & \,not OEF\, & not EF & Remark \\ \hline
\,CICY3\,& $7,868$ & 53 & 53 & \,$h^{1,1} > 4$ $\Rightarrow$ OEF\, \\
\hline 
CICY4 & \,$905,684$\, & 462 & \,not known\, & \,$h^{1,1} > 12$ $\Rightarrow$ OEF\, \\
\hline
\end{tabular}
\caption{
Obvious elliptic fibrations (OEF) and elliptic fibrations (EF) in the CICY3 and CICY4 datasets with $SU(3)$ and $SU(4)$ holonomies, respectively. 
\label{t:ell}}
\end{table}

Finally, it is worth noting that not every elliptic fibration admits a section. For instance, the bi-cubic $3$-fold,
$\def\arraystretch{1.15}\left[\begin{array}{c|ccc} 
\IP^2 &\, 3 \\ 
\IP^2 &\, 3\\ 
\end{array}\right]  \ , 
$
can be viewed as a fibration of elliptic curves over either of the two $\IP^2$ factors, and hence is elliptic. However, it does not admit a section but only a tri-section. Such elliptic fibrations with only a multi-section are sometimes called genus-one fibrations to emphasize the absence of a section. In this letter, however, an elliptic fibration refers to any fibration of elliptic curves over a base manifold, whether there exists a section or not.

\section{Machine-Learning Elliptic Fibrations}
As already discussed, elliptic fibrations are {\it typical} within the landscape of known Calabi-Yau manifolds. For instance, all CICY $3$-folds with $h^{1,1} > 4$ are elliptic; see Table~\ref{t:ell} for relevant statistics for both CICY3 and CICY4 datasets.
With small Hodge number, however, there are distinguished ones which are not elliptic. 
In particular, any cyclic manifold with $h^{1,1}=1$ (e.g. the Quintic $3$-fold) clearly cannot be elliptic. 

In the meantime most model-building approaches in string phenomenology have based on manifolds of {\it small} $h^{1,1}$ (or the mirror case of small $h^{2,1}$). This is because there are fewer parameters for them which enter into the effective potential \cite{Candelas:2016fdy,Candelas:2008wb,Candelas:2007ac,Constantin:2016xlj,Altman:2014bfa, Anderson:2007nc}. 
We will also focus especially on small-$h^{1,1}$ manifolds since those with {\it big enough} $h^{1,1}$ are essentially all elliptic; information on (non-)ellipticity can thus be thought of as being learnable mainly through such simpler manifolds. Specifically, the training set will consist of the ellipticity data for CICYs with $h^{1,1}$ up to a certain upper bound, as detailed in the following. 


\subsection{CICY Threefolds}
There is a total of 643 CICY $3$-folds with $h^{1,1} \leq 4$, 53 of which are not elliptically fibred \cite{Anderson:2017aux}.
Let us test whether standard machine-learning algorithms can distinguish these.
Because we have an {\it unbalanced} dataset, where those which {\it are} elliptic exceed those which {\it are not} by one order of magnitude, some enhancement is needed.  Luckily, any CICY configuration is equivalent to an arbitrary row and/or column permutation; this gives us a natural way to enhance the number 53.
We take 10 random permutations of the rows and the columns each for the 53 configuration matrices and 3 such permutations for the $643 - 53 = 590$.
This gives us a total of $10^2 \cdot 53 = 5300$ cases of 0 (non-elliptic) and $3^2 \cdot 590 = 5310$ cases of 1 (elliptic). Furthermore, we note that the maximal number of row and columns are respectively 6 and 7 for these configuration matrices, so we pad with zeros where necessary so that the each data-point is of the form
\begin{equation}
M_{6 \times 7} \longrightarrow 1 \mbox{ or } 0 \ .
\end{equation}
In fact, the above counting of enhanced dataset is only an upper bound due to symmetries in the configurations. Nevertheless, we end up with a total labelled dataset $\cD$ of size around 10K.

We now have a binary classification problem, for which we perform supervised machine-learning on a training set $\cT$ of size $x \%$ and validate on the {\it unseen} or complementary $(100 - x) \%$ validation set $\cV$.
By validation we mean to construct a $2 \times 2$ confusion matrix  $C$ where we record the number of false/true positives/negatives of the actual values compared to those predicted by the machine \cite{enc}:
\[
{\small
\begin{tabular}{cc|c|c|}
	\cline{3-4} 
	&  & \multicolumn{2}{c|}{{Actual}} \\ \cline{3-4}
	& &  True & False \\ \hline
	\multicolumn{1}{|c|}{{Predicted}} & True & True Positive
	($tp$) & False Positive ($fp$) \\  \cline{2-4}
	\multicolumn{1}{|c|}{{Classification}} & False & False Negative
	($fn$)& True Negative ($tn$)\\ \hline
\end{tabular}
}
\]
From $C$ we have certain ``goodness-of-prediction/fit''. There is the obvious ``precision'':
\begin{equation}
\text{Precision } := \frac{tp}{tp+fp}\,
\end{equation}
which is the percentage of correct prediction.
However, to avoid assigning too much weight to false positives, the standard accuracy measure to use is the Matthew's correlation coefficient $\phi$ for $C$:
\begin{equation}
\phi := \frac{tp \cdot tn - fp \cdot fn}{\sqrt{(tp+fp)(tp+fn)(tn+fp)(tn+fn)}}\,
\end{equation}
which is related to the chi-squared $\chi^2$ of $C$ as $|\phi| = \sqrt{\frac{\chi^2}{n}}$ where $n$ is the total validation size (i.e., sum of entries of $C$).
A value of $\phi = +1$ means completely accurate prediction, $-1$, complete anti-correlation and $0$, random guess \cite{phi}.

\begin{figure}[h!!!]
\begin{center}
Input $M_{6 \times 7} \rightarrow
\begin{array}{c}
\includegraphics[width=3.2cm]{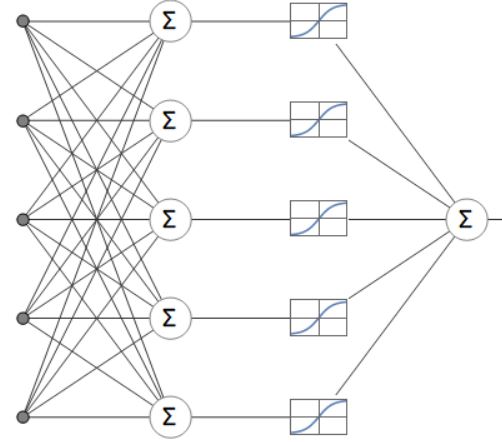}
\end{array}$
\begin{tabular}{l}
Output\\
0 or 1
\end{tabular}
\caption{\sf A graphical schematic of the neural network used (we have combined the element-wise Tanh and Sigmoid layers).
}
\label{f:NN}
\vspace{-21pt}
\end{center}
\end{figure}

For the actual AI, we choose a neural network which was found to be highly efficient in binary problems in computational geometry in \cite{He:2017aed,He:2017set}, viz., a 4-layer-perceptron consisting of (1) a fully connected linear layer $L_{5}$ taking the $6 \times 7$ matrix input by linear transformation to 5 nodes; (2) an element-wise layer of the sigmoid function $\sigma(z) = (1 - \exp(-z))^{-1}$; (3) an element-wise layer of the hyperbolic tangent function; (4) another fully-connected linear layer $L_5$ and (5) a summation layer of the 5 nodes into a single real output between 0 and 1, on which we then use integer round to return 0 or 1.
Schematically, the multi-layer perceptron looks like
\begin{equation}\label{NN}
\begin{array}{l}
    \fbox{\mbox{\begin{tabular}{l} INPUT =\\  $M_{6 \times 7}$ \end{tabular}}}
    \longrightarrow
    \fbox{$L_{5}$}
    \longrightarrow
    \fbox{$\sigma_{5}$}
    \longrightarrow
    \fbox{$\tanh_{5}$} \rightarrow \\
\quad
    \longrightarrow
    \fbox{$L_{5}$}
    \longrightarrow
    \fbox{Round($\Sigma_{1}$)}
    \longrightarrow
    \fbox{\mbox{\begin{tabular}{l} OUTPUT \\ = 0 or 1 \end{tabular} }}
\end{array}
\end{equation}
A graphic representation of the layers of the neural network is given in Fig.~\ref{f:NN}.

We cross-checked this neural network with other methods such as a classifier with an optimized mixture of decision trees and  regressions (implemented in the vanilla version of {\sf Classify[~]} function in \cite{math}) and find these also to perform to comparable precision.

\begin{figure}[h!!!]
\begin{center}
\includegraphics[width=8.3cm]{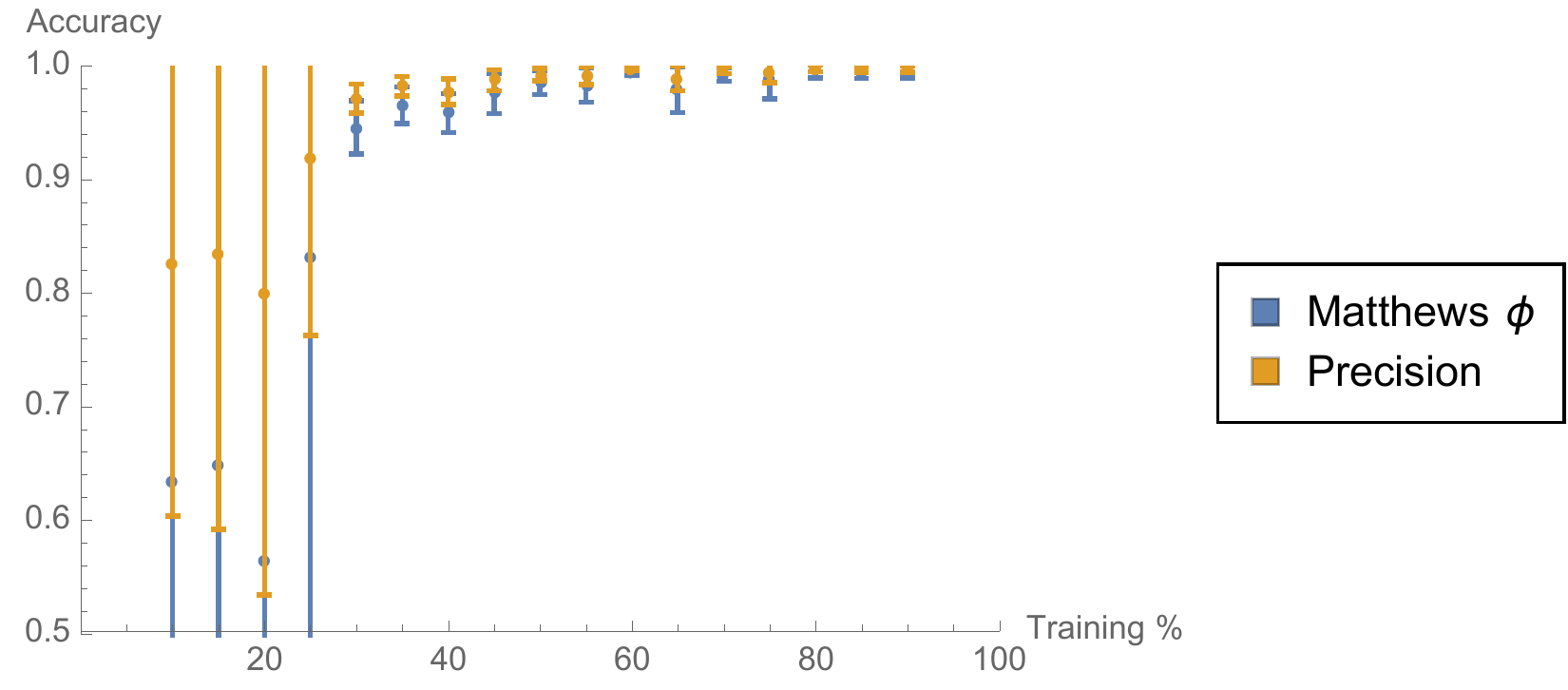}
\caption{\sf The learning curve for the (enhanced) 10K data for $h^{1,1}\leq 4$ CICY $3$-folds on distinguishing elliptic fibrations.
}
\label{f:tc-h11<=4}
\vspace{-21pt}
\end{center}
\end{figure}

We present the learning curve of the accuracy $\phi$ versus the percentage $x$ of seen training set in Fig.~\ref{f:tc-h11<=4}.
The error bars come from the fact that each training set at a given percentage is randomly chosen 10 times for statistical stability.
We see from the figure that the neural network works very well with respect to both of the accuracy measures. Initially, up to about 25\% there is a lot of fluctuation because too few data has been seen to learn anything meaningful.  However, starting from around 30\%, we see that both naive precision (\% of agreement of 0s and 1s) and Matthews' $\phi$ approach 1.
What is remarkable is that even at a relatively low percentage of seen data, we are achieving high accuracy.
For example, having seen only about 30\% of the total 10K of data of what is an elliptic fibration, the neural network correctly predicts the unseen 70\% of cases to over 99\% accuracy.
Importantly, the computation for each training takes only a few seconds, a vast improvement over any brute-force check of whether a manifold admits an elliptic fibration structure.
In summary, determining whether a CICY is elliptically fibred indeed seems to be a machine-learnable problem.

\subsection{Control Case}
It was argued in \cite{He:2017aed,He:2017set} that computational problems in algebraic geometry tend to fall into the machine-learnable category because essentially {\it any problem} therein boils down to a finite steps of finding kernels and cokernels of integer matrices, a task in which the likes of neural networks excel (problems in number theory, in contrast, seem much less amenable to AI).
The machine certainly knows nothing about manifolds or bundle fibrations, all it is doing is to catalogue the positive/negative cases together and to manage to spot a pattern not immediately obvious to the human eye (or, for that matter, to the standard methods of algebraic geometry).
One might question then whether if one divides any random property into any random subsets, the machine might pick up something, whereby making any specific queries such as elliptic fibration structures rather ineffectual.

It is therefore a good and important thing to test a ``control case'' as follows.
Suppose of the $643$ CICY $3$-fold configurations, we select 53 {\it arbitrarily} and {\it randomly} and assign a property as ``0'' while all other cases we will call ``1'' in complete imitation to the above problem and repeat the machine-learning procedure.
If elliptic fibration is truly {\it not a random property}, then the AI should perform poorly in accuracy to this artificial problem.

\begin{figure}[h!!!]
\begin{center}
\includegraphics[width=8.3cm]{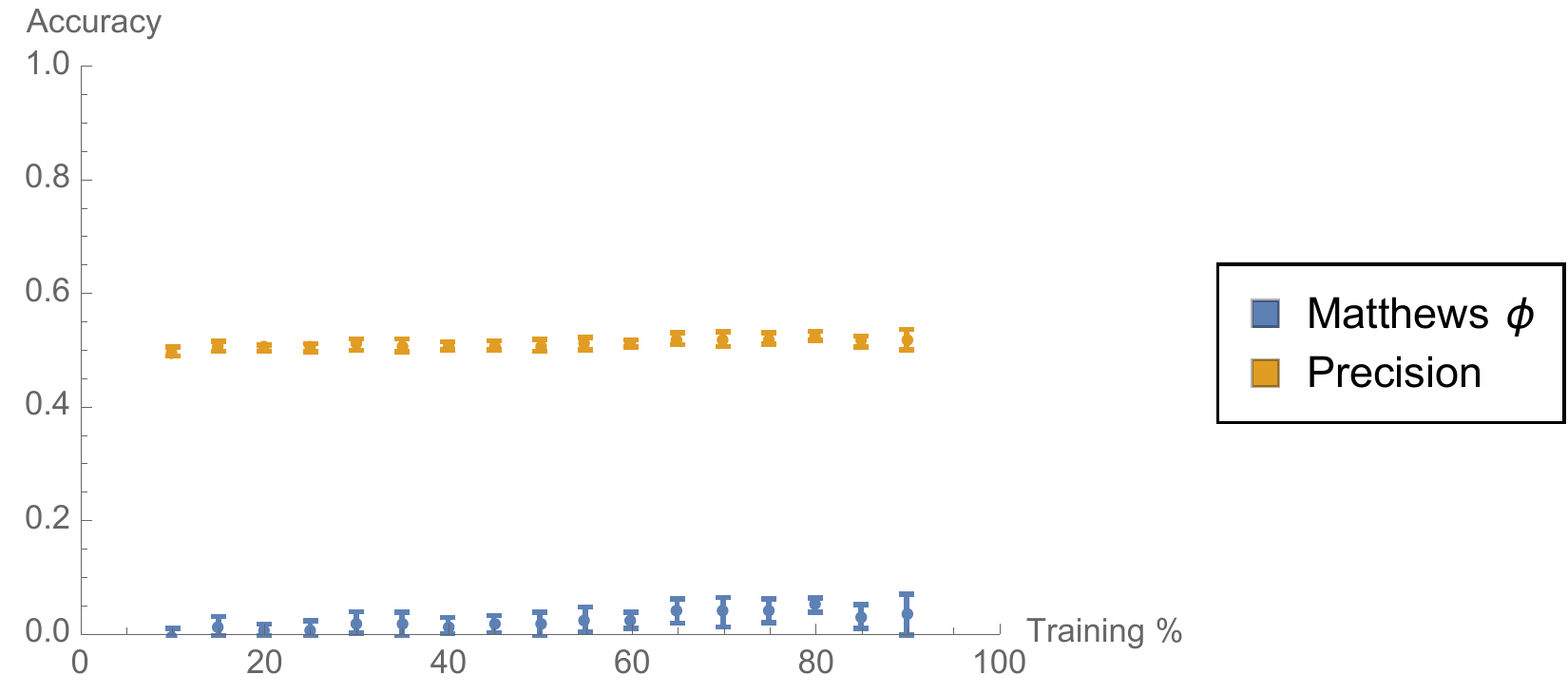}
\caption{\sf The learning curve for the (enhanced) 10K data for $h^{1,1}\leq 4$ CICY $3$-folds on a control set of a randomly chosen property.
}
\label{f:tc-h11<=4control}
\vspace{-21pt}
\end{center}
\end{figure}

The learning curve is presented in Fig.~\ref{f:tc-h11<=4control}. It is reassuring that the results are in agreement with complete randomness.
The naive precision stays at 50\%, consistent with a random guess and subsequently, the $\phi$ coefficient remains essentially at 0.
The control case thus shows that a random assignment of a property does not and should not allow the machine to spot any inherent pattern.

\subsection{CICY Fourfolds}
Since the machine knows nothing about algebraic geometry, there is no reason to stop at CICY 3-folds.
The next dataset, which is even more significant to realistic F-theory vacua, is the Calabi-Yau 4-folds arising as complete intersections in products of projective spaces. As reviewed in Section~\ref{CICY-EF}, $921,497$ CICY $4$-fold configurations were classified in~\cite{Gray:2013mja} and their OEF structures in~\cite{Gray:2014fla}; see Table~\ref{t:ell} for the relevant statistics. 

\begin{figure}[h!!!]
\begin{center}
\includegraphics[width=8.3cm]{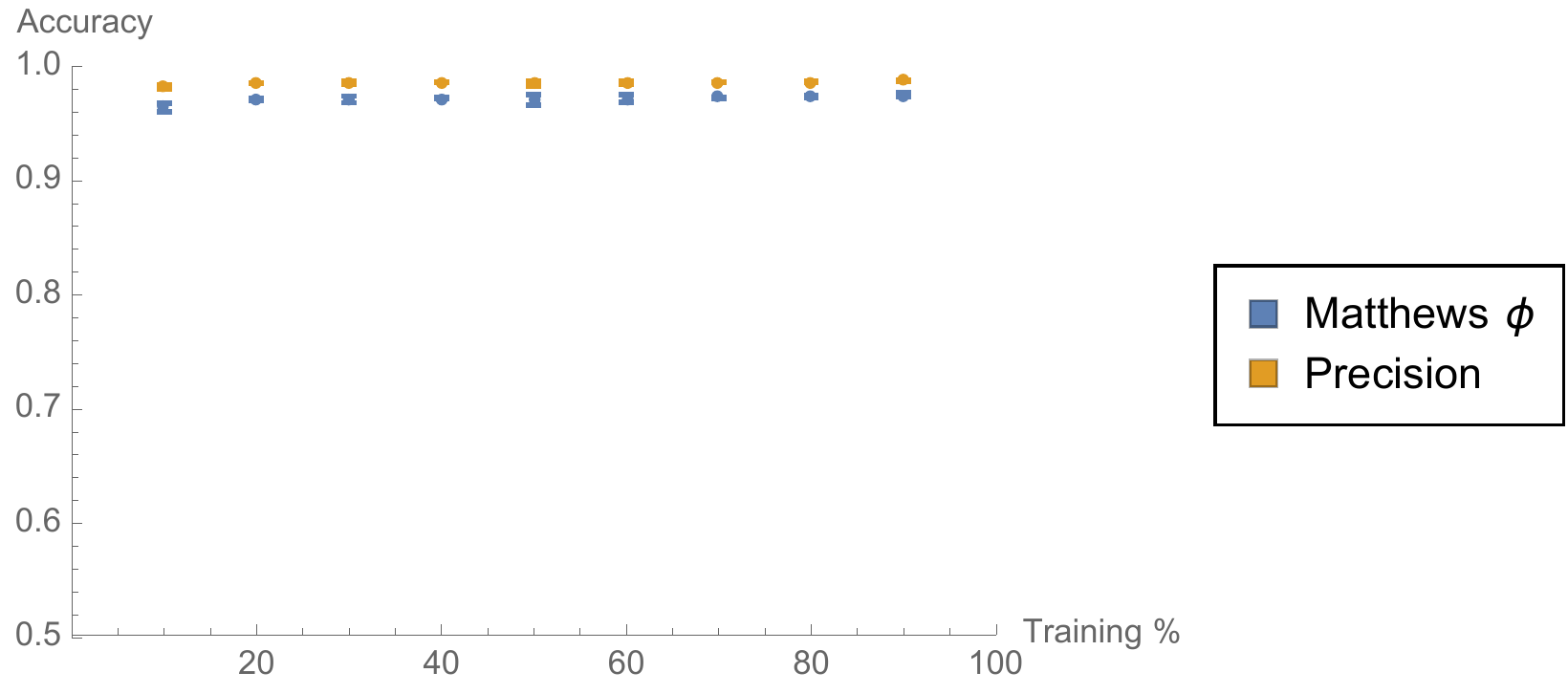}
\caption{\sf The learning curve for the (enhanced) 1.54M data for $h^{1,1} \leq 12$ CICY 4-folds on distinguishing elliptic fibrations.
}
\label{f:tcCICY4}
\vspace{-21pt}
\end{center}
\end{figure}

Up to $h^{1,1} = 12$, there is now a total of 767,642 configurations, 767,180 of which admit elliptic fibrations.
Here, everything is a $12 \times 16$ integer configuration matrix, upon padding with zeros where necessary as in the 3-fold case.
Again, due to the paucity of non-elliptic fibrations, we enhance data by permutations: we randomly permute the non-elliptic ones by row/columns $140^2 = 19600$ times and the elliptic ones by $1^2 = 1$, thereby obtaining a more balanced $\sim$ 767K for 1 (elliptic) and 774K for 0 (non-elliptic) upon removal of redundancies.
The learning curve for the CICY 4-folds is presented in Fig.~\ref{f:tcCICY4}. The (small) error bars are due to the fact that each training set at a given percentage $x=10\%, \dots, 90 \%$ is randomly chosen $5$ times for stability. 
Impressive is the fact that even better than the $3$-fold case, the neural network correctly predicts the unseen cases to over $96\%$ accuracy already at $x=10\%$.

We performed a similar control case for the CICY 4-folds, i.e., taking a random sample of $767642 - 767180=462$ cases and declaring them to have property ``0'' and the remainder, property ``1''.  After data enhancement by permutation, more on the ``0'' than the ``1'', we find a training curve identical to Fig.~\ref{f:tc-h11<=4control}. That is, when there is no pattern, the neural network will give a completely random guess with Matthew's phi equaling 0.

Interestingly, if we train, on the CICY 3-fold set, on which ones are elliptically fibred and attempt to predict which CICY 4-folds are elliptic, the network consistently  (over repeated trials) predicts that all CICY 4-folds are elliptically fibred.
This is actually not too far from the truth since non-elliptic fibration is comparatively rare.
However, working within the {\it same} dataset (using 3-folds to predict about 3-folds, and 4-folds, about 4-folds) actually picks out which of the specific rare cases that are not elliptic.

~\\

\section{Conclusions and Outlook}
In this letter, we have demonstrated that distinguishing whether a complete intersection Calabi-Yau manifold is elliptic is a problem well suited for machine-learning.
The fact that we have used a large dataset of specific Calabi-Yau manifolds is only for the purposes of being concrete; it is expected that the general problem of identifying fibration structure for arbitrary varieties should also be addressable by the likes of neural networks.

On another matter, distinguishing the elliptic fibrations with a section from those without one is in general a complicated task in algebraic geometry~\cite{Anderson:2016ler}. 
It will be interesting to see if machine-learning techniques can be employed into such a task. Yet another interesting aspect of elliptic CICYs is that they in general admit more than one fibration structures~\cite{Anderson:2016cdu}, which was fully analyzed for CICY 3-folds~\cite{Anderson:2017aux}. Machine-learning aspects of such an enumeration problem, we leave to future investigation. 

In some sense, the key question of \cite{He:2017aed,He:2017set} is ``what class of problems in mathematics and computation can be machine-learned''.
In particular, it was found that typical problems in algebraic geometry, such as cohomology calculation or toric variety combinatorics, seem to be better suited for a neural network than for traditional methods.
After all, as far as the AI is concerned, it only knows a trainable set of inputs (configuration matrices of manifolds or bundles, etc.) and outputs (ranks of cohomology groups, dimensions of loci, yes/no answer of existence of fibrations), perfectly adapted for supervised learning. From this, the AI finds some optimal arrangement, by regression, classifiers, neural networks, decision trees, etc., in order to guess the answer in general.

This is at once mysterious and invigorating. As is well-known, computational problems in algebraic geometry ultimately suffers from the exponentially expensive steps of finding Groebner bases and syzygies, because of the high degree multi-variate polynomials involved.
Very quickly, one would find the desired problem intractable even with the most advanced algorithms and computers.
That the neural network is able to find the correct answer in a matter of seconds (precisely because it knows nothing of the mathematics) suggests that there is some underlying method/structure which is yet to be discovered.  In this sense, our ``control experiment'', not emphasized in previous studies, is important. We found that for a random assignment of fictitious property, the AI gives completely random guesses ($\phi \simeq 0$, precision $\simeq 50\%$). This suggests that there truly exists an underlying pattern to meaningful properties such as whether a manifold is an elliptic fibration.

\end{document}